\documentclass{article}

\usepackage{PRIMEarxiv}

\usepackage[utf8]{inputenc} 
\usepackage[T1]{fontenc}    
\usepackage{hyperref}       
\usepackage{url}            
\usepackage{booktabs}       
\usepackage{amsfonts}       
\usepackage{nicefrac}       
\usepackage{microtype}      
\usepackage{lipsum}
\usepackage{fancyhdr}       
\usepackage{graphicx}       
\graphicspath{{media/}}     
\usepackage{tabularray}
\usepackage{pdflscape}
\usepackage[longtable]{multirow}
\usepackage{longtable}
\title{Intellectual Property Blockchain Odyssey: Navigating Challenges and Seizing Opportunities}

\author{
  Rabia Bajwa \\
  MCTI Program  \\
  University of Guelph \\
  Guelph\\
  \texttt{Rbajwa01@uoguelph.ca} \\
   \And
  Farah Tasnur Meem \\
  MCTI Program  \\
  University of Guelph \\
  Guelph\\
  \texttt{ftasnurm@uoguelph.ca} \\
}

\begin{document}
\maketitle

\begin{abstract}
  This paper investigates the evolving relationship between Intellectual Property Rights (IPRs) protection and blockchain technology. We conducted a comprehensive literature review, supplemented by case study analyses and research paper reviews, in order to understand the scope and implications of blockchain in relation to intellectual property rights. Our study demonstrates how applying blockchain technology for IPR could revolutionize transparency, security, and operational efficiency. It also identifies the primary challenges and openings in this area. We provide an extensive framework for integrating blockchain technology with intellectual property rights, as well as other technical components (some of which already exist or are resolved by blockchain; some might need attention), drawing on current research and best practices. This framework has the potential to give a new perspective in a structured manner for the intellectual property landscape by providing 360-degree coverage across different layers of operation.
\end{abstract}

\keywords{Smart Contract, Intellectual Property Rights (IPR), Patents, Trademarks, Trade Secrets, Copyrights}

\section{Introduction}

\subsection{Background and context of the research}
\begin{figure}[ht]
\centering
  \includegraphics[width=0.5\columnwidth]{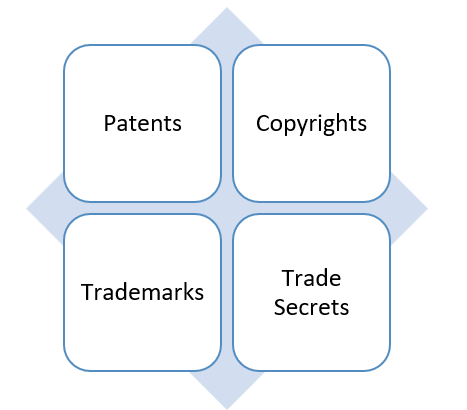}
  \caption{Intellectual Property Rights (IPR) Types}~\label{fig:figure2}
  \label{fig:fiqure2}
\end{figure}

Intellectual property (IP) encompasses human creations, which are fundamentally intangible, as well as their creative works, like ideas and knowledge. The types of intellectual property rights that are internationally recognized are trade secrets, copyrights, patents, and trademarks \ref{fig:figure2}.

Due to IP's intangible nature, which makes it harder to protect than more conventional, tangible assets like cash, real estate, and merchandise, IP protection is more difficult. As the virtual economy grows quickly, disputes pertaining to infringement of intellectual property rights—including those related to photography, film, copyrights, trademarks, creativity, and other areas—occur frequently. Knowledge and creativity are becoming more digitally accessible, which facilitates content replication while making it harder and more expensive to claim originality. Potential applications of blockchain technology in intellectual property include establishing and enforcing IP agreements, licenses, or exclusive distribution networks through smart contracts; managing digital rights; registering and clearing IP rights; providing proof of creatorship and provenance authentication; and sending real-time payments to IP owners. \cite{r15} 

Regarding patents, the true advantage of blockchain technology is its ability to create an unchangeable record of information with a tamper-proof code that offers solid proof of facts regarding the life cycle of an invention. Nevertheless, in contrast to copyrights, any new creation will still need to be patented with the appropriate authorization; otherwise, anybody else can use it freely and not face any legal repercussions for imitation or claim. 

\textbf{Blockchain-based applications for IPR management}
Any kind of intellectual property, including trade secrets, utility models, patents, and know-how, could be the subject of a blockchain application because the goal of each use case will determine the free definition of the transaction's object. (Public) IP registers, such as the European or German patent and trademark registers, could be one area of use. For instance, the parties could directly enter assignments of IP rights, licenses, or patent pledges into a blockchain-based IP register (with the relevant software API provided), which would cut down on the time and expense of office proceedings. Furthermore, it might make the register more trustworthy with regard to the actual legal circumstances, guaranteeing that the register is maintained more current 
Any work results (such as freelancers' or collaboration partners' contributions) or, more accurately, the intellectual property's corresponding "digital fingerprint" (hash value) can be stored in the blockchain within the context of any development agreement. This allows the contracting parties and any other third parties to confirm the IPR's creation at a specific time (Proof-of-Existence). Such evidence may be especially crucial if the relevant legal system, like German copyright law, does not have a registration mechanism that grants an individual absolute rights over a specific subject.

As such, the World Intellectual Property Organization convened its inaugural conference on blockchain applications in the domain of intellectual property datasets in 2019.\cite{r15} The main objective of WIPO's "Blockchain Task Force" aims to "explore the possibility of using blockchain technology in the processes of providing IP rights protection, processing information about IP objects and their use" \cite{r17}. Within that framework, WIPO's "Blockchain Whitepaper Project" seeks to produce a whitepaper that will investigate opportunities and obstacles in more detail, pinpoint possible applications, and formulate suggestions regarding interoperability and governance concerning the intellectual property ecosystem.\cite{r18}

Blockchain may also be useful for managing access authorization and license granting in relation to smart contracts. For instance, the blockchain would only validate the user's payment once access to digital online content \cite{r19} (such as music, videos, photos, and other documents) was granted \cite{r19}. Similar to this, in the case of licensing relationships, the smart contract/blockchain could monitor the issuance of licenses and/or sublicenses as well as the timely payment of royalties (which could be determined by other data or the quantity of licenses obtained).

\subsection{Issues in existing IPR} 

\begin{itemize}
    \item \textbf{Ethical concerns:}The control of necessary technologies is a matter of ethics when intellectual property rights result in monopolies. The problem stems from the ethical quandary of giving one organization substantial control over an endeavor that may benefit many people, as this could lead to misuse of authority and oppressive policies. \newline It is a challenge to address the unequal distribution of benefits resulting from intellectual property rights. There is a concern that the existing system may disadvantage people and smaller entities in favor of larger corporations or those with greater resources.\end{itemize}

\begin{itemize}
    \item \textbf{Deciding on the proper scope of patent protection:} Choosing the patents' scope is a challenging task. Patents that are overly broad can hinder innovation by prohibiting third parties from developing upon already developed technologies. They might not offer enough of an incentive for innovation if they are overly restricted. Finding a balance that safeguards inventors and encourages additional innovation is the difficult part. \end{itemize}
    
    \begin{itemize}
    \item \textbf{Handling patent trolls:} Innovators face a great deal of difficulty when dealing with patent trolls, who purchase patents solely to file infringement lawsuits without ever producing anything. They can discourage investment in novel ideas by sowing fear and uncertainty and depleting resources through litigation. \end{itemize}

    \begin{itemize}
    \item \textbf{Intellectual property rights may deter researchers from sharing their work:} When researchers fear losing control over their work, they may be reluctant to share their findings due to IP rights. The problem is that this can make research environments more static and isolated while also impeding collaborative efforts, which are essential for scientific progress.\end{itemize}

\subsection{Statement of the problem}  \emph{
The study tackles a major problem: the efficient use of blockchain technology for intellectual property rights (IPR) management. Blockchain technology is able to track and verify the owner of an idea or creation, but it is still in its infancy when it comes to managing the finer points of intellectual property rights (IPR), such as contract interpretation or verifying that research is conducted in accordance with established objectives. Additionally, a balance between the rights of inventors and creators and the technology's capacity to distribute digital content widely are openly. Solving this problem is essential to adjusting IPR management to the changing digital environment influenced by blockchain developments.}

\subsection{Objectives of the study} 

\begin{itemize}
    \item
\textbf{To close the gap between potential theory and real-world application:} Examine the factors influencing blockchain's sluggish uptake in IPR management.Examine what is needed in a blockchain system to manage intellectual property rights (IPR) efficiently.\end{itemize}

\begin{itemize}
    \item \textbf{To create a customized framework for the incorporation of blockchain technology into IPR:} Draft a system hypothetical design that addresses the unique requirements of intellectual property rights management. Ensure that contractual subtleties and legal complexities are covered by the framework.\end{itemize}

\begin{itemize}
    \item \textbf{To enable the use of blockchain to enable a more reliable and efficient IPR system:}
Present a well-thought-out plan for using blockchain technology to improve IPR management's transparency, equity, and effectiveness. Strive to make the system for monitoring, distributing, and upholding intellectual property rights better overall. \end{itemize}

\begin{itemize}
    \item \textbf{
 To offer suggestions to interested parties regarding the use of blockchain in IPR: }
Provide legal entities, users, and creators with advice on navigating the blockchain IPR system. Make recommendations for best practices for blockchain governance and implementation in the field of intellectual property. \end{itemize}


\section{Terminologies} 
 \textbf{Tokens:} Tokens are cryptoassets that operate on an existing blockchain network instead of their own. Whilst tokens can also be used in a similar fashion to coins, they are often created to fulfil different purposes to coins, for example to raise funds or give access to particular services. Some examples of tokens include Shiba Inu, Tether, and Basic Attention Token. \cite{r9} \newline 
      \textbf{“Smart” IP rights:} \newline
    The potential to use blockchain technology for the management of IP rights is vast. Recording IP rights in a distributed ledger rather than a traditional database could effectively turn them into “smart IP rights”\cite{r10}\newline
     \textbf{Patent:} A patent is a legal right that allows innovators to have exclusive control over their inventions for a set period of time, usually 20 years. It forbids others from creating, using, or selling the patented invention without the consent of the patent holder. please give me broad information in 1 paragraph for the patent along with this content.  \cite{r12} \newline
    \textbf{Copyrights:} \newline
    Copyright safeguards creators of literary, artistic, and musical works by granting exclusive rights to reproduce, distribute, perform, and display their creations. This protection lasts for the creator's lifetime plus a specific number of years, fostering creativity by providing a financial incentive. \cite{r13}\newline
     \textbf{Trademark:} \newline A trademark, such as a logo or name, serves as a unique identifier distinguishing products or services. Registered trademarks provide legal protection, preventing unauthorized use and reinforcing brand recognition in the competitive market.\cite{r13} \newline
    \textbf{Trade secrets:} \newline A trade secret is confidential business information that gives an advantage in the marketplace. This information, such as formulas, processes, or customer lists, is not made public and is legally protected. Businesses must take reasonable steps to keep information confidential in order to maintain trade secret status. \cite{r14}
 \newline
\section{Literature Review} 
Gonenc et al.\cite{ref2} Intellectual property law and practice in the blockchain realm. Here, he examines the legal implications of blockchain technology, as well as its potential impact on intellectual property (IP) law. It investigates the possibilities for using blockchain in IP registration, management, and enforcement, recommending solutions for IP offices, customs procedures, and rights management efficiency. The conclusion includes recommendations for furthering blockchain technology and expanding its integration into various services and registration channels.
Eva et al.\cite{r4} The chapter examines intellectual property (IP) and its complicated environment in relation to blockchain technology. It examines the difficulties governments, lawmakers, and intellectual property owners encounter when navigating this complex field, where billions of dollars are at risk and a new industrial revolution could occur. The conversation covers a broad overview of blockchain technology, how it is used to manage intellectual property rights, and the larger context of IP rights in the blockchain industry. It places special emphasis on the economic ramifications and the conflict that exists between those who are at the forefront of technology and those who want unrestricted access to its advantages.
Huang et al.\cite{r5} addresses the difficulties in upholding copyrights in the contemporary legal and technological environment, emphasizing two primary problems: establishing authorship or ownership and carrying out copyright transactions. Blockchain is suggested as a solution for document preservation and authentication, especially when combined with IPFS. But possible software updates give rise to worries. Although smart contracts and the Ethereum Blockchain are seen as a way to lower transaction costs, there are legal issues to be resolved, such as party identification, contract modifications, code-based justifications, and dispute resolution. Concerns regarding copyright infringement are raised by the anonymity of blockchain users. In order to provide effective copyright protection in this dynamic environment, the copyright legal system must keep up with the advancements in blockchain and related technologies.\newline
Junyao and Shenling \cite{r6} This study examines how blockchain technology and intellectual property interact, evaluating both scholarly and industrial uses. It explores the tamper-resistant and decentralized features of blockchain, demonstrating how it can be used to solve problems related to intellectual property management. The review covers a wide range of topics, such as copyright protection, trademark management, and patent registration, and it highlights scholarly contributions and successful case studies. The amalgamation of scholarly discoveries and pragmatic applications serves as the foundation for determining forthcoming pathways in the persistent assimilation of blockchain technology within the domain of intellectual property. 
Yue shi and Qingqing \cite{r7} emphasizes how well the decentralization, tamper-resistance, and transaction anonymity of blockchain technology can solve traditional intellectual property problems, such as the challenges associated with acquiring electronic evidence and the high expense and inadequate compensation associated with copyright protection. Blockchain ensures safe and unchangeable recording, improving intellectual property protection, by distributing information among all nodes, removing middlemen, and employing encryption algorithms. The paper explores the technical foundation of blockchain, describing its features, and discusses how optimization techniques like proof of stake and cross-chain can help overcome the present constraints on transaction throughput, latency, and resource usage. In the end, it looks at current business issues and provides predictions for how blockchain technology will advance intellectual property development in the future. 
Kensuke and Marcus \cite{r8} This chapter conducts a thorough examination of the application of blockchain technology in intellectual property management, filling a gap in critical discussions regarding its feasibility. Despite widespread discussions in media and industry about the potential of blockchain in this domain, the paper takes a nuanced approach by critically assessing potential limitations and proposing tentative solutions. The focus is on two key perspectives: the operational aspects and the implementation of blockchain technology for intellectual property management. The chapter argues that beyond the technical attributes often emphasized, the incentive design, rooted in the original Bitcoin proposal, plays a crucial role in achieving genuinely decentralized and disintermediated intellectual property management.

\section{Research Methodology}
The research specifically focuses on the problem of using blockchain technology to address the present issues with managing intellectual property rights (IPRs). Blockchain provides answers to problems related to IPR rights enforcement, transparency, and proof of ownership, but there is a big gap in its real-world implementation in this field. This disconnect is caused by a lack of knowledge about the capabilities of blockchain technology and the lack of customized frameworks that can connect it to the complex needs of IPR management. With this research, we hope to shed light on this intersection and create a workable framework for integrating blockchain technology into IPR systems.

\begin{itemize}
    \item Gathering information on how blockchain is perceived in general and within IP community
\end{itemize}
\begin{itemize}
    \item Analyzing implications of blockchain applications in IP space
\end{itemize}

\begin{itemize}
    \item Identifying and analyzing the specific challenges and inefficiencies currently faced in IPR management that blockchain technology could address
\end{itemize}
\begin{itemize}
    \item Developing a detailed, actionable framework for integrating blockchain into the IPR ecosystem, addressing legal, technical, and operational aspects.
\end{itemize}

\textbf{Gathering information on how blockchain is perceived:}
Firstly we focused on understanding the general perception of blockchain technology and its specific perception within the intellectual property (IP). By conducting literature reviews on blockchain technology to grasp its broader applications, strengths, and weaknesses we aim to uncover any preconceived notions, concerns, or expectations that may influence the integration of blockchain into intellectual property rights management.  \newline
\textbf{Analyzing implications of blockchain applications in IP space:} 
Analyzing the implications of blockchain applications in the intellectual property (IP) space involves a comprehensive examination of how the integration of blockchain technology could impact various aspects of IP management. In this phase  delved into existing use cases and applications where blockchain technology has been deployed within the realm of intellectual property. By scrutinizing these real-world examples, we identified the potential strengths and weaknesses of blockchain applications in addressing specific challenges faced in IP management. The analysis encompassed a review of how blockchain enhances transparency, facilitates proof of ownership, and contributes to the enforcement of intellectual property rights. \newline
\textbf{Identifying and analyzing the specific challenges and inefficiencies currently faced in IPR management that blockchain technology could address:}
By examining scholarly articles, reports, and reputable online platforms, the research aims to identify recurring issues such as complex and time-consuming proof of ownership(PoW) processes, lack of transparency in IP transactions, and difficulties in enforcing rights. The absence of a centralized and tamper-proof ledger exacerbates challenges related to fraudulent claims and disputes. The analysis considers how blockchain's decentralized and transparent nature could potentially mitigate these challenges by providing an immutable record of ownership, enhancing transparency in transactions, and streamlining the enforcement of intellectual property rights. This evidence-based approach allows for a comprehensive understanding of the specific pain points within IPR management that could be effectively addressed through blockchain technology.

\section{Proposed Framework}
\begin{figure}[ht]
    \centering
    \includegraphics[width=0.8\textwidth]{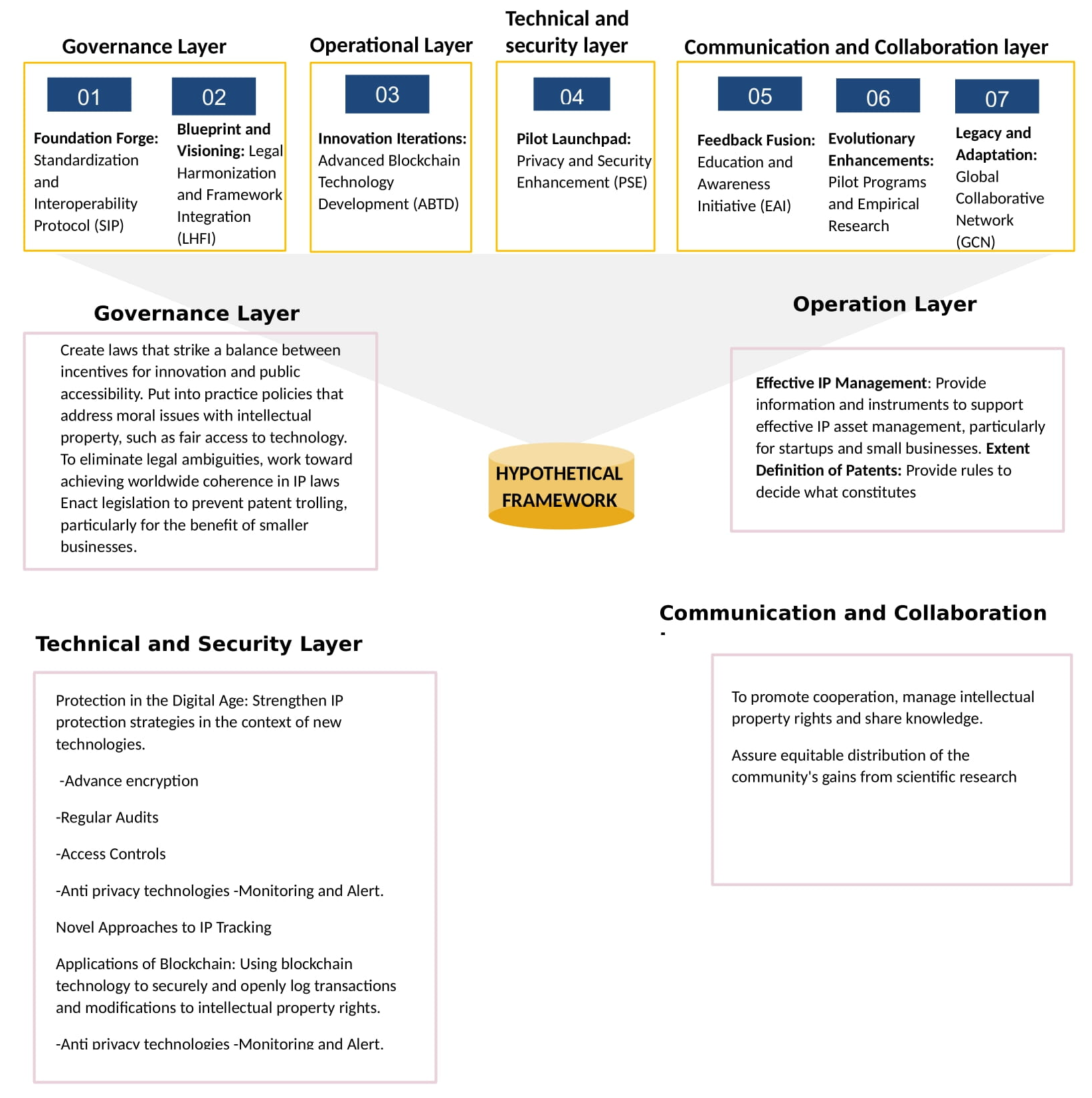}
    \caption{Proposed framework}
    \label{fig:framework}
\end{figure}

We tried to propose a detailed and actionable framework [Figure 02] for the seamless integration of blockchain into the intellectual property rights (IPR) ecosystem in the our research. This entails a multifaceted approach that addresses legal, technical, and operational issues. On the legal front, we intend to outline the regulatory considerations and legal frameworks that must be in place to ensure compliance and legitimacy in the use of blockchain for intellectual property management. We intend to provide a clear roadmap for the implementation of blockchain technology, taking into account factors such as interoperability with existing systems, data privacy, and security. Our framework will delve into the operational aspects of incorporating blockchain into day-to-day IPR processes, such as user education, system maintenance, and security. 
\cite{ref2}
\ref{fig:framework}
\subsection{5.1 Governance Layer}
\textbf{Foundation Forge: Standardization and Interoperability Protocol (SIP)
} \newline
\textit{Standardization}: Developing a unified global patent/IP system: Creating a single, worldwide patent and intellectual property system: This calls for the application of blockchain technology to unify national patent systems. A global patent/IP system based on blockchain technology could be the foundation for improving IP management effectiveness and spurring innovation. \newline
\textit{Interoperability Protocol}: Laws and governments everywhere realizing blockchain's potential: this is the stage at which the idea of incorporating blockchain technology into legal frameworks is conceived. Developing a unified legal framework for intellectual property management would include accepting blockchain as admissible evidence and utilizing it as a proof of evidence. \newline
\textit{Blueprint and Visioning:Legal Harmonization and Framework Integration (LHF)}
This element aims to support the decentralized and global nature of blockchain by harmonizing the legal systems of various nations or areas. It seeks to lessen legal disputes and make international IP rights transactions easier.Framework Integration: This refers to combining the capabilities of blockchain technology with standardized legal frameworks to create a legal-tech interface that can effectively manage intellectual property rights (IPR) issues.

\subsection{Operational Layer}
\textit{Innovation Interation: Advanced Blockchain Technology Development (ABTD)}:
This refers to the continuous innovation and advancement of blockchain technology, which is customized to meet IP management requirements. For example, scaling, speed, and privacy features are improved to meet the demands of intellectual property rights transactions.
\textit{Pilot Launchpad: Privacy and Security Enhancement (PSE)}
Privacy: Ensuring the confidentiality of sensitive data pertaining to IPR transactions. The use of private or permissioned blockchains and privacy-enhancing technologies like zk-SNARKs are two examples of how blockchain technology can be used for privacy.
Security Enhancement: fortifying the blockchain against possible flaws that might allow intellectual property to be stolen, leaked, or corrupted.

\subsection{Technical and Security  Layer}
\textit{Blockchain for digital asset security:} Because of its immutability, security, and transparency, blockchain technology is ideal for securing digital assets, and this idea would be embodied by the central framework. This involves managing digital assets' entire lifecycle and utilizing blockchain as proof in court.
\textit{Versioning and indexing:} Providing each file a distinct fingerprint and enabling versioning and indexing are among the fundamental features of the central framework.Intellectual property (IP) such as patents, research and publications, copyrights, and other digital assets necessitate multiple versions over time. It justifies the creation of technological advancements that allow for the unified archiving of various digital asset iterations throughout their entire lifecycle. Blockchain technology can be applied to systems so that users can link all versions of their digital assets together using its ledger technology, and it can even be used for asset lifecycle management from beginning to end.
By making an invention publicly known and thereby creating prior art for it, defensive publications serve as a countermeasure against patent applications. Every file has a distinct fingerprint, duplicates are eliminated, versioning is enabled by the platform, network nodes can select what content to host, the database is searchable and indexed, and blockchain could function as a platform for defensive publications.

\begin{figure}[ht]
    \centering
    \includegraphics[width=0.8\textwidth]{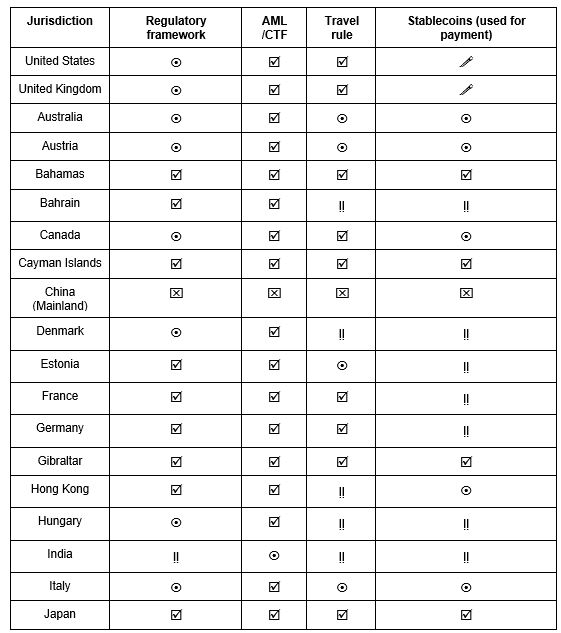}
    \caption{Comparison of AML Regulations, Travel Rule, and Stablecoin Usage Across Jurisdictions}
    \label{fig:Table}
\end{figure}

Other technical controls that should be incorporated in all block chain technology platforms(some of the platforms already incorporate these controls):
\textit{Advanced Encryption:}
Asymmetric cryptography may be used in advanced encryption to secure transactions in the context of blockchain and intellectual property rights. Every party would possess a set of keys, a public key for encrypting transactions and a private key for decrypting them, so that only authorized parties could view the data.
\textit{Frequent Examinations:}
Because all transactions on a blockchain are recorded on a public ledger, they are auditable by nature. Examining the blockchain for the correctness and legitimacy of IPR transactions would be part of routine audits. These audits could be automated by smart contracts to guarantee adherence to IPR laws.
\textit{Anticipation and Warning of Anti-Privacy Technologies:}
In the context of blockchain, anti-privacy technology might include programs that keep an eye on the ledger for any unauthorized use or access to intellectual property rights. If intellectual property owners' rights are being violated, smart contracts can be configured to send out alerts.
\textit{Maintaining Integrity:}
Strong integrity protection is offered by the blockchain's immutability. The inability to change an IPR-related transaction once it has been added to the blockchain helps to guard against fraud and guarantees the authenticity of the IP record.
Using Digital Watermarks:
By logging the watermark details for every asset on the blockchain, blockchain can facilitate digital watermarking. By doing this, the watermark would become permanently recorded, which is essential for establishing ownership and identifying unauthorized use.
\textit{Digital Rights Management System (DRM):}
Smart contracts that carry out license agreements automatically can be used to improve DRM on blockchain. To prevent unauthorized reproduction, a smart contract might, for instance, mandate that a digital piece of art can only be printed once.
\textit{GenAI Protection: }
One possible approach to safeguarding against intellectual property infringement on the internet is to use machine learning algorithms to keep an eye out for IP infringements. To identify unauthorized use or duplication, the AI would compare online content with the IP stored on the blockchain.
\textit{Tamperproof:}
The tamperproof ledger of blockchain ensures that data entered is unchangeable. Since it offers a reliable and unchangeable record of IP creation, transfer, and licensing, this is essential for intellectual property rights (IPRs). \ref{fig:Table}
\subsection{Communication and Collaboration Layer}

\textit{Feedback Fusion: Education and Awareness Initiative (EAI)}
Informing interested parties about the ways in which blockchain technology can be applied to defensive publication, transparency, and the avoidance of patenting already-disclosed inventions.
\textit{Evolutionary Enhancements: Pilot Programs and Empirical Research (PPER)}
This approach calls for the gradual, iterative development of blockchain applications for IPR management. It entails conducting empirical research to compile data and insights from the real world and implementing controlled pilot programs to test blockchain in particular IPR sectors. With a focus on ongoing improvement based on real-world application and research findings, this approach guarantees the integration of blockchain technology in IPR management in a practical and efficient manner.

\textit{Legacy and Adaptation: Global Collaborative Network (GCN)}
• \textit{Legacy:} Assessing the current IP management systems and drawing lessons from both their achievements and shortcomings is part of this process.

\

• \textit{Adaptation: }This refers to modifying smart contracts to manage intricate licensing agreements in order to tailor blockchain technology to the particular requirements of the IPR context.
The establishment of a global network of intellectual property rights stakeholders aims to promote information exchange, agreement on intellectual property rules, and cooperative creation of optimal methodologies.
• \textit{"Smart" R\&D agreements and ownership management of intellectual property:} This layer is in line with the concept of employing blockchain for "smart" R\&D agreements when IP is licensed and new IPRs are created. In order to handle ownership distribution and project IPR licensing, partners must work together and communicate with one another."Smart" research and development ("R\&D") agreements, in which the partners license each other's intellectual property to produce new intellectual property rights, could be another example of a use case. The ownership distribution of (new) IPR and the (unilateral/mutual) licensing of project IPR can both be managed with blockchain technology. Additionally, funding could become available if it is verified that particular project milestones—like the successful development of an initial prototype—have been reached.

\section{Future work and conclusion}
Promoting regulatory compliance and setting industry standards for blockchain-based IPR systems is a critical need for future research. It is imperative to engage in constructive dialogue with regulatory authorities, legal specialists, and industry stakeholders in order to impact the development of industry standards and compliance protocols that can guarantee the legal soundness and international recognition of blockchain-based intellectual property solutions. Through continuous interaction, the legal landscapes of various jurisdictions will be navigated and an environment favorable to the broad adoption of blockchain technology for IPR management will be shaped.Some of the future possible considerations or potential use cases are as follows:

Better Counterfeiting Measures: Blockchain technology can provide manufacturers and brands with an impenetrable way to track products from manufacturing to retail. As a result, there may be a considerable decrease in counterfeit goods, safeguarding consumer confidence and intellectual property.

Worldwide IPR Licensing Platform: A blockchain-based platform might make it easier to license intellectual property internationally. It would offer a clear, effective international marketplace for the purchase, sale, or licensing of intellectual property rights.

Digital Rights Management (DRM): Blockchain technology has the potential to provide more transparent and resilient digital rights management (DRM) systems. It can monitor and control how digital content is used and distributed, which helps stop unapproved distribution.

Monetization of User-Generated Content: Blockchain technology can open up new revenue streams for social media and other platforms' user-generated content. It can guarantee that content producers receive just compensation for their valuable contributions to the platform.

Cross-Industry IP Clearances: Blockchain can streamline the process of obtaining and utilizing intellectual property rights across various industries, lowering legal hurdles and expediting production schedules, in industries like media and advertising.


\bibliographystyle{unsrt}
\bibliography{bibs_file.bib}

\end{document}